\documentclass[aps,amsmath,amssymb,preprintnumbers,groupedaddress,twocolumn,floatfix]{revtex4}            

\usepackage{graphicx} 
\usepackage{pstricks}

\newcommand{\nc}{\newcommand}
\nc{\beq}{\begin{equation}}
\nc{\eeq}{\end{equation}}
\nc{\bea}{\begin{eqnarray}}
\nc{\eea}{\end{eqnarray}}
\def\ov{\overline}
\def\IR{\mathbb{R}}

\begin{document}


\preprint{MPP-2008-147}

\title{Gauge Coupling Unification in F-theory GUT Models}

\author{Ralph Blumenhagen} \email{blumenha@mppmu.mpg.de}

\affiliation{Max-Planck-Institut f\"ur Physik, F\"ohringer Ring 6, 80805 M\"unchen, Germany}


\begin{abstract}
We investigate gauge coupling unification for F-theory
respectively Type IIB orientifold constructions
of $SU(5)$  GUT  theories with  gauge symmetry breaking via 
non-trivial hypercharge  flux. 
This flux has the non-trivial
effect that it splits the values of the three MSSM gauge couplings
at the string scale, thus  potentially spoiling the celebrated one-loop gauge coupling
unification.  
It is shown how F-theory can evade this problem in a natural way.
\end{abstract}



\maketitle




\section{Introduction}

The unification of the three gauge couplings of the strong
and electroweak interactions at a scale $M_X=2.1\cdot 10^{16}\,$GeV
in the minimal supersymmetric Standard Model (MSSM) is 
the strongest argument for the existence
of a unifying grand unification  at this high scale 
\cite{Ellis:1990zq,Amaldi:1991cn,Giunti:1991ta,Langacker:1991an}.
The minimal simple  gauge groups containing 
$SU(3)_c\times SU(2)_w\times U(1)_Y$ are $SU(5)$ and
$SO(10)$. Besides the MSSM particles, these contain
extra ones which have to receive a mass in the process
of breaking the GUT gauge group. For a Georgi-Glashow
$SU(5)$ GUT theory \cite{Georgi:1974sy}, there exist in particular the 
$X$ and $Y$ gauge bosons and in addition a vector-like
pair of  Higgs triplets  
$({\bf 3}, {\bf 1})_{-{2\over 3}}+({\bf \ov 3}, {\bf 1})_{{2\over 3}}$
which combine with the weak Higgs doublet  into the   
${\bf 5}_H+{\bf \ov 5}_H$ representation.

Combining the idea of a unification of gauge couplings
with the unification of the gravitational interaction  seems
to be quite natural, as the  GUT scale and Planck-scale 
$M_{pl}=2.4\cdot 10^{18}\,$GeV are not so far apart.
Concrete examples  for such a unification naturally arise from
String Theory, where a compactification of for instance
the $E_8\times E_8$ ten-dimensional heterotic string
on a Calabi-Yau threefold  ${\cal X}$ can lead to a 
four-dimensional effective field theory
with ${\cal N}=1$
supersymmetry and gauge group $SU(5)$ or $SO(10)$.
Since in most these string models there does not exist a light
adjoint Higgs field, one has to implement  an alternative  
mechanism by which the GUT gauge group is broken. 
In most cases this is done by turning
on discrete Wilson lines supported on homologically non-trivial
1-cycles in the internal compact manifold. 

Very recently, in the context of F-theory resp. Type IIB orientifold
compactifications a different possibility  has been
made very concrete \cite{Donagi:2008ca,Beasley:2008dc}.
Here the $SU(5)$ gauge group
is supported on a stack of $D7$ branes wrapping a surface  
in the internal Calabi-Yau manifold ${\cal X}$.
The three generations of chiral matter in the ${\bf 10}+{\bf \ov 5}$
representation are localised on curves where the GUT brane intersects
other branes. The same  happens for  the Higgs field  
${\bf 5}_H+{\bf \ov 5}_H$ which is also localised on such a curve.
The GUT surface is chosen to be rigid, i.e. it  is a  del-Pezzo surface,
so that there are no candidate $SU(5)$ adjoint Higgs fields.   
Moreover, a del-Pezzo surface does not admit any discrete Wilson line.
 
Similar to earlier considerations for the heterotic string
\cite{Witten:1985bz,Blumenhagen:2006ux,Tatar:2008zj},
this leaves  a third  possibility to break the $SU(5)$ gauge group
\cite{Beasley:2008kw}.
One can turn on an internal  gauge flux $\ov{f}_Y$ 
with values in the $U(1)_Y$. One can describe this flux
as a connection on some line bundle ${\cal L}_Y$, whose
structure group breaks the  $SU(5)$  to the SM gauge group. Usually, this
would lead to St\"uckelberg masses for the four-dimensional hypercharge 
gauge field. However, if the $U(1)_Y$ flux is localised
on a non-trivial two-cycle in the del-Pezzo, which is 
trivial when considered as a two-cycle in Calabi-Yau,
the mass mixing with the axions can be avoided
and the final gauge group is indeed
$SU(3)_c\times SU(2)_w\times U(1)_Y$ \cite{Buican:2006sn}.
We refer the 
reader to \cite{Donagi:2008ca,Beasley:2008dc,Beasley:2008kw,Donagi:2008kj,
Marsano:2008py} 
for more details on such local F-theory models 
or to \cite{Plauschinn:2008yd, Blumenhagen:2008at} 
for the realisation on compact Type IIB orientifolds.
A couple of phenomenological features of such models, like
Yukawa textures \cite{Font:2008id,Heckman:2008qa} and
supersymmetry breaking \cite{Marsano:2008jq},
have been investigated recently.

\begin{figure}[ht]
\begin{center} 
\includegraphics[width=0.4\textwidth]{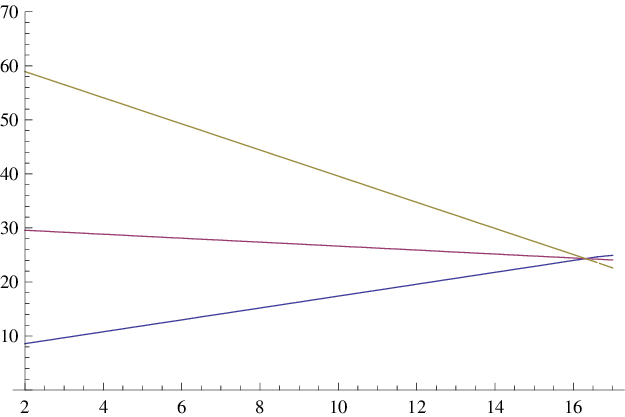}
\begin{picture}(0,0)
  \put(0,8){${\rm log}_{10}(\mu)$}
   \put(-193,120){${3\over 5}\alpha_Y^{-1}$}
   \put(-193,68){$\alpha_w^{-1}$}
   \put(-193,32){$\alpha_s^{-1}$}
\end{picture}
\hspace*{15pt}
\end{center} 
\vspace{-15pt}
\caption{{\small 
One-loop running of gauge couplings  for   MSSM light matter 
using the parameters
$M_Z=91.18$GeV, $\alpha_s(M_Z)=0.1172$, $\alpha(M_Z)={1\over 127.934}$
and $\sin^2\theta_w(M_Z)=0.23113$ .}\label{figa}}
\vspace*{-5pt}
\end{figure}
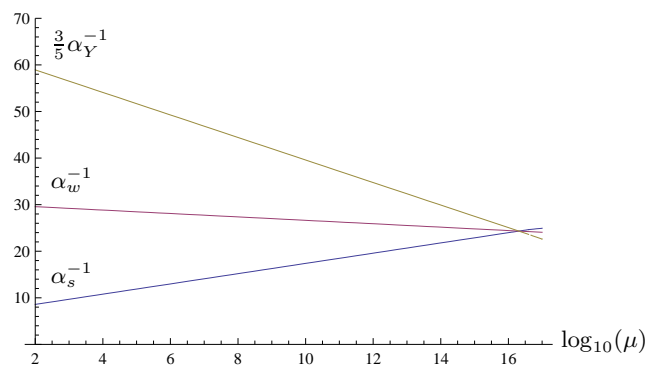

It is the aim of this letter to analyse the important issue
of gauge coupling unification for such F-theory/IIB orientifold
$SU(5)$ GUT models in more detail.
Employing the formalism and notation put forward in \cite{Blumenhagen:2008at},
all computations are carried out in the Type IIB
orientifold framework though  they should carry over mutatis mutandis to
the more general F-theory framework. 
The main new issue here is the presence of the $U(1)_Y$ flux, 
which at first sight causes a serious problem, in that
it  splits the values of the
three MSSM gauge couplings at the string resp. unification scale.
This  might spoil  the beautiful unification of these couplings
shown in figure \ref{figa}.

\section{Gauge couplings in F-theory GUT}

Let us assume that in  either a local or a global model
we have identified a del-Pezzo surface $D_a={\rm dP}_r$ on which  
we can wrap five D7-branes supporting  a $U(5)$ gauge group. 
This surface is embedded into the Calabi-Yau as
$\iota: D_a\to {\cal X}$. 
As shown in \cite{Blumenhagen:2008at}, via the Freed-Witten gauge flux
quantisation condition, the fact that the del-Pezzo is non-{\it Spin} implies
that these D7-branes must support a non-trivial line bundle
${\cal L}_a$.  In the following we will choose 
this line bundle to come from a restriction
of a line bundle on ${\cal X}$, i.e. ${\cal L}_a=\iota^*(L_a)$.

The matter fields ${\bf 10}+{\bf\overline 5}$ 
result from intersections of this $SU(5)$ brane
 with its orientifold image respectively  a second single brane
wrapping a divisor $D_b$ which supports  a line bundle ${\cal L}_b$.
In  F-theory  these matter fields are described as
enhancements of the $SU(5)$ degeneration of the elliptic fiber
of the fourfold to $SO(10)$ resp.
$SU(6)$ along  certain matter curves.

Now, one breaks the $SU(5)$ GUT theory by 
 turning on a non-trivial $U(1)_Y$ flux ${\cal L}_Y$
supported on a two-cycle on the del-Pezzo surface
which is however trivial in the Calabi-Yau, i.e. $\iota_*({\cal L}_Y)=0$.
Clearly, if the breaking of the $SU(5)$ is such that below
the breaking scale one has precisely the MSSM matter,
the running of the gauge couplings is such that 
at the one-loop level they unify at $M_X=2.1\cdot 10^{16}$GeV. 
The most natural scenario is that one identifies
the GUT scale with the string scale.

Let us consider the tree-level gauge couplings at the
string scale. The $SU(5)$ gauge kinetic function 
$ f_{SU(5)}= {4\pi\over g^2_X} + i \Theta$ is simply given by
\bea
       f_{SU(5)} = \tau_a ={1\over 2\, g_s\ell_s^4}\int_{D_a} J\wedge J +i  \int_{D_a}
       C_4\; ,
\eea
where $g_s=e^\varphi$ denotes the string coupling constant and
${\rm Vol}(D_a)={1\over 2 }\int_{D_a} J\wedge J$ is the volume
of the del-Pezzo surface $D_a$.
However, the presence of the line bundles ${\cal L}_a$, 
${\cal L}_Y$ generates subleading terms, which can be 
computed by  
dimensionally reducing the Chern-Simons action of the D7-brane
wrapping the del-Pezzo surface $D_a$
\bea
\label{chernsim}
          S_{CS}=\mu_7 \,  \int_{D_a\times \IR^{1,3}} C_0\wedge {\rm tr} (F^4)
       \; .
\eea
In our case the overall flux $F$ has the following expansion
\bea
\label{fluxexpans}
    F&=&\sum_{a=1}^8  F^a_{SU(3)} 
        \left(\begin{matrix}  
             \lambda_a/2 & 0 \\
             0 & 0  
          \end{matrix}\right) 
        +\sum_{i=1}^3  F^i_{SU(2)} 
        \left(\begin{matrix}  
             0 & 0 \\
             0 & \sigma_i/2  
          \end{matrix}\right)+ \nonumber\\
     &&   {\textstyle {1\over 6}}\, F_Y 
          \left(\begin{matrix}  
             -2_{3\times 3} & 0 \\
             0 & 3_{2\times 2} 
          \end{matrix}\right)+\\
     &&\!\!\!\!\!\! \Bigl(\overline{f}_a +{\textstyle {2\over 5}}\, \overline{f}_Y \Bigr)
          \left(\begin{matrix}  
             1_{3\times 3} & 0 \\
             0 & 1_{2\times 2} 
          \end{matrix}\right) +
        {\textstyle {1\over 5}}\, \overline{f}_Y
          \left(\begin{matrix}  
             -2_{3\times 3} & 0 \\
             0 & 3_{2\times 2} 
          \end{matrix}\right)\, ,\nonumber
\eea
where $\lambda_a$ denote the eight traceless Gell-Man matrices and
$\sigma_i$ the three traceless Pauli matrices.
The capital letters $F_G$ denote the four-dimensional gauge fields
and the small letters $\ov f$ the internal background fluxes.   
Now, inserting the expansion (\ref{fluxexpans}) into the Chern-Simons term
(\ref{chernsim}) and extracting the $F\wedge F$ terms, we eventually
find the three tree level gauge kinetic functions
\footnote{These formulas differ from those derived in \cite{Donagi:2008kj}, 
   which is due  to the fact that here a different 
      embedding, eq. (\ref{fluxexpans}), of the fractional line bundle
 ${\cal L}_Y$ into $U(5)$ has been used.}
\bea
\label{gaugekins}
      f_{SU(3)}&=& \tau_a -{1\over 2}\, S\, \int_{D_a} c^2_1({\cal L}_a) \nonumber \\ 
      f_{SU(2)}&=& \tau_a -{1\over 2}\, S\, \int_{D_a} \bigl(\, c^2_1({\cal L}_a)
      +c^2_1({\cal L}_Y)\, \bigr)\\
     {3\over 5}\,  f_{U(1)_Y}&=& \tau_a -{1\over 2}\, S\, \int_{D_a}  
      \bigl(\, c^2_1({\cal L}_a) +{3\over 5}  c^2_1({\cal L}_Y)\, \bigr) \; , 
\nonumber
\eea
where we used  ${\cal L}_a=\iota^*(L_a)$ 
and $S=e^{-\varphi}+i C_0$ denotes the axio-dilaton field
\footnote{Twisting ${\cal L}_a=\iota^*(L_a)$ by a ``trivial'' line bundle, i.e.
${\cal L}_a\to {\cal L}_a\otimes {\cal R}_a$ with $\iota_*({\cal R}_a)=0$, 
eq. (\ref{gaugekins}) changes such that $\int c^2_1({\cal L}_Y)\to
\int [\,  c^2_1 ({\cal L}_Y)+  2\, c_1 ({\cal L}_Y)\,  c_1 ({\cal R}_a)\, ]$.
Note that for $e_Y=c_1({\cal L}_Y)$ and $e_a=c_1({\cal R}_a)$ being  roots of $E_r$ 
with $e_Y\cdot e_a=1$ this correction vanishes, thus  leading
to ordinary gauge coupling unification.   }.
As usual  in string theory, these couplings receive
one-loop threshold corrections at order $M_{KK}\simeq O(M_s)$ 
with $M_{KK}\simeq (1/{\rm Vol}(D_a))^{1\over 4}$, whose
effect  we ignore in the following at leading order.

For models without  light exotics 
$({\bf 3},{\bf 2})_{5\over 3}+({\bf \ov 3},{\bf 2})_{-{5\over 3}}$ originating
from non-vanishing cohomology groups $H^*(D_a,{\cal L}_Y)$,
we have to choose a line bundle ${\cal L}_Y$ supported  on the $E_r$ sublattice
of $H^2({\rm dP}_r,\mathbb Z)$ with  $\int c^2_1({\cal L}_Y)=-2$.
In other words, $c_1({\cal L}_Y)$ has to correspond to a root of
$E_r$. We expect the masses of the lightest such states to be of  order 
$M_{KK}$.
Turning on this $SU(5)$ symmetry breaking flux, for finite $g_s$ 
the gauge couplings
at the string scale $M_s$ do not unify any longer.

One might be tempted that this could explain the 4\% deviation
from MSSM gauge coupling unification at the 2-loop level \cite{Ibe:2003ys}.
However, for F-theory/IIB orientifold  GUTs the order of the 
gauge couplings at the string scale is
\bea
\label{rela} 
    {1\over \alpha_s(M_s)}< {3\over 5\, \alpha_Y(M_s)} <{1\over
      \alpha_w(M_s)}\; ,
\eea
which never occurs in \cite{Ibe:2003ys} for any value of the scale $\mu$. 
Therefore, we seem to have a serious problem with gauge coupling unification
in this class of  GUT models.
\vspace{0.2cm}

\section{Higgs triplet threshold} 

As shown in the last section, at the string scale we find the
relation
\bea
\label{relb} 
      \Delta_{13}={3\over 5}\, \Delta_{23}
\eea
for $\Delta_{13}={3\over 5}{\alpha}^{-1}_Y-{\alpha}^{-1}_s$
and $\Delta_{23}={\alpha}^{-1}_w-{\alpha}^{-1}_s$.
Intriguingly, this can also be written as 

\bea
\label{relc}
        {1\over \alpha_Y(M_s)}={1\over \alpha_w(M_s)} + {2\over 3\,
          \alpha_s(M_s)}\; ,
\eea
a relation which has already  appeared for gauge couplings
in more general Standard Model like four stack intersecting D6-brane 
models \cite{Blumenhagen:2003jy}(see also \cite{Ibanez:1998rf}).

It is clear that, in order to have any chance that running up the low-energy
couplings to the string scale,  they satisfy relation 
(\ref{rela}) and (\ref{relb}), there must exist a new threshold in between.
These new charged states must contribute to the running
such that it creates a region for $\mu$, where the order
of the gauge couplings is like in (\ref{rela}).

Recall that prior to GUT symmetry breaking, there were the
Higgs fields ${\bf 5}_H+ {\bf \overline 5}_H$ and that
the $U(1)_Y$ bundle was chosen such that the 
doublets remain massless and the triplets become massive, i.e.
\bea
\label{higgscohoms}
 H^*(C, {\cal L}^{-1}_a\otimes {\cal L}_b\otimes K^{1\over 2}_C )&=&(0,0) \\
 H^*(C, {\cal L}^{-1}_a\otimes {\cal L}_b \otimes {\cal L}^{-1}_Y \otimes
K^{1\over 2}_C )&=&(1,1)\; . \nonumber
\eea
Here $C$ denotes the curve $C\subset D_a$ supporting  the Higgs fields
and $K_C$ its canonical line bundle.
In the case that the Higgs field is localised on a curve
of genus one, the mass of the triplet is determined
by a non-trivial Wilson line originating from the reduction
of the line bundles on the genus one curve. 
These states have masses smaller than the Kaluza-Klein scale $M_{KK}$,
which we assume to be (slightly) smaller than the order of the string scale.
In \cite{Beasley:2008kw} a second option was discussed, where the two Higgs
fields  are localised on different curves, so that 
via (\ref{higgscohoms}) the masses of the triplets come
from pairings of ${\bf 3}_H, {\bf \ov 3}_{H}$ with other fields
respectively.
This allows to suppress dimension five proton decay operators.

Therefore, these colour triplets $({\bf 3},{\bf 1})_{-{2\over 3}} +
({\bf\overline 3},{\bf 1})_{{2\over 3}}$ are the distinguished
natural candidates to change the running
of the gauge couplings at a scale $M_{3\overline 3}<M_X$
and we treat their mass scale as a parameter 
$1{\rm TeV}< M_{3\overline 3}< M_X$ for the following analysis. 

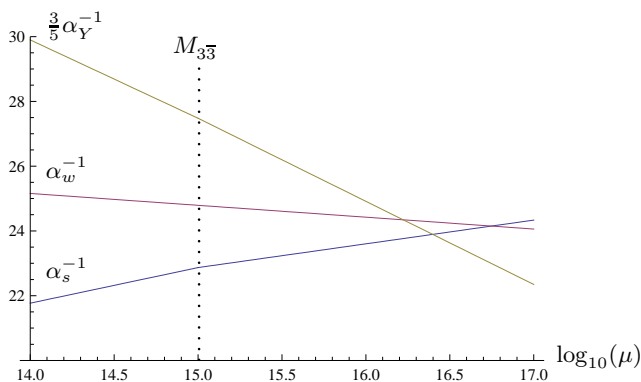
\begin{figure}[ht]
\begin{center} 
\includegraphics[width=0.4\textwidth]{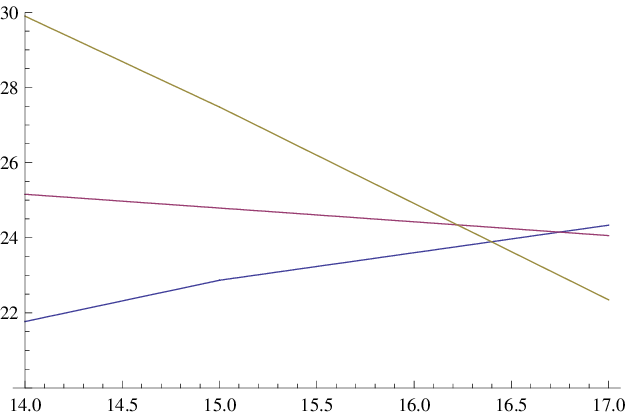}
\begin{picture}(0,0)
  \put(0,8){${\rm log}_{10}(\mu)$}
   \put(-193,133){${3\over 5}\alpha_Y^{-1}$}
   \put(-193,78){$\alpha_w^{-1}$}
   \put(-193,40){$\alpha_s^{-1}$}
   \put(-145,125){$M_{3\overline 3}$}
   \psline[linewidth=1pt, linestyle=dotted]{-}(-4.75,0.3)(-4.75,4.2)
\end{picture}
\hspace*{15pt}
\end{center} 
\vspace{-15pt}
\caption{{\small 
One-loop running of the MSSM matter with Higgs triplet
threshold at $M_{3\overline 3}=10^{15}$GeV.} \label{figb}}
\vspace*{-5pt}
\end{figure}

Above this new threshold the MSSM beta-function coefficients 
change according to
\bea
  && (b_3,b_2,b_1) =(3,-1,-11)\to  \\
  && \phantom{aaaaaaaaaaaaaaaa} \to(\tilde b_3,\tilde b_2,\tilde b_1)= 
(2,-1,-{35\over 3})\; . \nonumber
\eea
Choosing for instance $M_{3\overline 3}=10^{15}\,$GeV, the running
around the GUT scale changes as shown in figure \ref{figb}.

 \vspace{0.7cm}

\vspace{0.4cm}

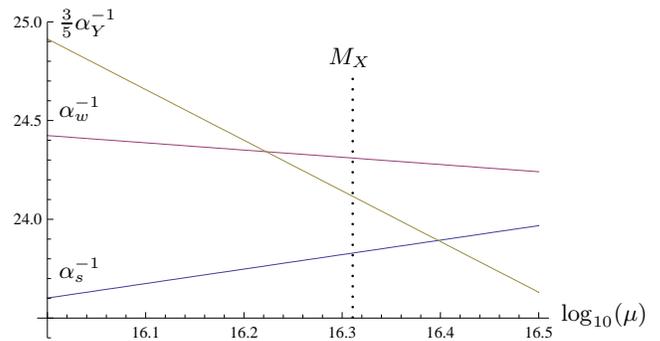
\begin{figure}[ht]
\begin{center} 
\includegraphics[width=0.4\textwidth]{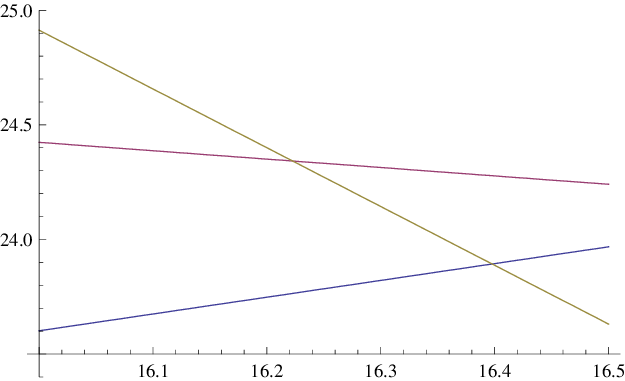}
\begin{picture}(0,0)
  \put(0,8){${\rm log}_{10}(\mu)$}
   \put(-190,118){${3\over 5}\alpha_Y^{-1}$}
   \put(-190,85){$\alpha_w^{-1}$}
   \put(-190,24){$\alpha_s^{-1}$}
    \put(-88,105){$M_{X}$}
   \psline[linewidth=1pt, linestyle=dotted]{-}(-2.77,0.3)(-2.77,3.5)
  \end{picture}
\hspace*{15pt}
\end{center} 
\vspace{-15pt}
\caption{{\small 
One-loop running of the gauge couplings beyond
the Higgs triplet threshold .}\label{figc}}
\vspace*{-5pt}
\end{figure}

It is obvious that now there exist a region where the
order of the gauge couplings is as in (\ref{rela}).
Zooming in into this region we get figure \ref{figc}.

Now, in the region $16.2< \log_{10} (\mu) < 16.4$ there must exist 
a point where precisely the  relation (\ref{relb}) holds.
Indeed, numerically we find that this happens
at 
\bea
               M_X=2.1\cdot 10^{16}\,{\rm GeV}
\eea
which quite remarkably is  the value of the usual GUT scale.

\section{Gauge coupling F-unification}

For analysing what happens around the GUT scale, we have to
look more closely at  the running of the gauge couplings between the
threshold $M_{3\overline 3}$ and the GUT scale.
In fact, the running of the three gauge couplings at scales 
$\mu>M_{3\overline 3}$ can be written as
\bea
\label{running}
    {\textstyle {1\over  \alpha_s(\mu)} }&=& {\textstyle {1\over \alpha_s(M_Z)} + {b_3\over 2\pi} 
                  \log\left({\mu\over M_Z}\right)+
          {\tilde b_3- b_3 \over 2\pi} \log\left( 
        {\mu\over M_{3\overline 3} } \right)} \nonumber \\
  {\textstyle  {1\over  \alpha_w(\mu)} }&=& {\textstyle {\sin^2 \theta_w\over \alpha (M_Z)} + 
       {b_2\over 2\pi} \log\left({\mu\over M_Z}\right) }\\
     {\textstyle   {1\over   \alpha_Y(\mu)} }&=& 
    {\textstyle   {\cos^2 \theta_w\over \alpha (M_Z)} 
             + {b_1\over 2\pi} 
                  \log\left({\mu\over M_Z}\right) 
    +{\tilde b_1- b_1 \over 2\pi} \log\left( 
        {\mu\over M_{3\overline 3} } \right) }\; . \nonumber 
\eea
Now requiring that the F-theory GUT relation (\ref{relb}) holds 
at $\mu=M_{X}=M_s$, leads to the simple relation
\bea
   \left( (\tilde b_1- b_1) -  {2\over 3} (\tilde b_3- b_3 )\right)\log\left( 
        {M_X\over M_{3\overline 3} } \right) =0 
\eea
which is satisfied for any number of 
Higgs triplets and any  threshold
scale $M_{3\overline 3}<M_X$.
 
Let us summarise the main observations made in this letter:

\begin{itemize}
\item The breaking of the $SU(5)$ GUT via a 
non-trivial $U(1)_Y$ flux 
leads to a splitting of the three MSSM gauge couplings at the
string/GUT scale
\bea
\label{onerelation}
        {1\over \alpha_Y(M_s)}={1\over \alpha_w(M_s)} + {2\over 3\alpha_s(M_s)}\; ,
\eea
thus spoiling the usual gauge  coupling unification.

\item
If the Higgs triplet  $({\bf 3}, {\bf 1})_{-{2\over 3}}+({\bf \ov 3}, {\bf
  1})_{{2\over 3}}$ is lighter than 
the GUT scale, this threshold changes the one-loop running of the
gauge couplings such that they  satisfy this F-theory GUT relation
at $M_X$ independent of the value of the threshold scale.

\item  In this case, from eq. (\ref{gaugekins}), eq. (\ref{running})
and  $c_1({\cal L}_Y)$ being a root of $E_r$, 
one can derive the following two additional relations
with $\alpha_X^{-1}\simeq 24$
\bea
\label{stringrelat}
       {1\over g_s}={1\over 2\pi} \log\left( {M_X\over M_{3\overline 3}}
            \right),\quad\
       {M_{KK}\over M_s} \simeq \left({ \alpha_X\over g_s}\right)^{1\over 4}
\eea

\end{itemize}

If the appearance of baryon number violating
dim$=5$ operators $QQQL$ forces
us to choose $M_{3\overline 3}$ of the order of 
$M_X$, we get $g_s>1$, i.e. we are driven  
to the strong coupling regime, where 
F-theory is expected to be the appropriate  description and where
for $g_s\gg 1$ all corrections in (\ref{gaugekins}) become negligible small.
However, if there exists  a mechanism to suppress these dangerous
$QQQL$ operators, $M_{3\overline 3}$ can be significantly
smaller than $M_X$, while  gauge coupling F-unification still holds.
For $M_{3\overline 3}<10^{13}\,$ GeV we even get 
$g_s<1$ with  $M_{KK}=O(M_X)$
\footnote{I thank D. L\"ust and T. Weigand 
for helpful discussions.}. 

Recall that in field theory GUTs  one gets two relations
among the gauge coupling constants at the GUT scale
leading to one prediction for the couplings at
the weak scale.
In the F-theory case, one has instead only one direct relation 
among the gauge couplings (\ref{onerelation}) at the string scale,
which already suffices to fix  the GUT scale.
In addition one finds one relation among the string parameters
$g_s$ and $M_{3\overline 3}$. If for a concrete string model
one has prior knowledge that  this relation holds, 
we also have one prediction among the gauge couplings at $M_w$.
In the more general scheme with $g_s$ and
$M_{3\overline 3}$ treated as adjustable  parameters
F-theory is less predictive than a field theory GUT.

\section{Comments}

From the observation made in this letter one can 
draw two not unrelated conclusions.
First, it shows the robustness
of the $SU(5)$ gauge breaking mechanism by a $U(1)_Y$ flux.
The  shift in the string scale gauge couplings can 
be reconciled  precisely by the running of the 
Higgs triplet below the string scale.
Second, in all $SU(5)$ GUT models with the mass of the  Higgs triplet 
significantly below the GUT scale, unification at $M_X=2.1\cdot 10^{16}\,$GeV
requires the F-theory like split (\ref{relc}) in the GUT scale gauge couplings.

Finally, it is amusing that by writing equation (\ref{stringrelat})
as $M_{3\overline 3}=M_s\, \exp(-2\pi \Re (S) )$ it looks,
as if the Higgs triplet gets its dominant mass from 
a $D(-1)$-brane instanton.\\

\emph{Acknowledgements:}
I am very much  indebted to Costas Bachas for reminding me
to revisit the issue of gauge coupling unification in
F-theory GUTs and to Dieter L\"ust and Timo Weigand 
for very useful discussions.
Moreover,  I thank  Michael K\"unkel and
Sebastian Moster for further discussions.


\end{document}